\documentclass[11pt,a4paper]{article}

\usepackage{jinstpub}
\usepackage{amssymb}
\usepackage{amsmath}
\usepackage{booktabs}

\title{\boldmath Magnetoresistance in copper at high frequency and high magnetic fields} 

\author[a]{S. Ahn,}
\author[b,1]{S.W. Youn,\note{Corresponding author.}}
\author[a,b]{J. Yoo,}
\author[a]{D.L. Kim,}
\author[b]{J. Jeong,}
\author[a]{M. Ahn,}
\author[b]{J. Kim,}
\author[b]{D. Lee,}
\author[a]{J. Lee,}
\author[a]{T. Seong,}
\author[a,b]{Y. K. Semertzidis}

\affiliation[a]{Department of Physics, Korea Advanced Institute of Science and Technology (KAIST), \\Daejeon 34141, Republic of Korea}
\affiliation[b]{Center for Axion and Precision Physics Research, Institute for Basic Science, \\Daejeon 34047, Republic of Korea}

\emailAdd{swyoun@ibs.re.kr}

\abstract{
In halo dark matter axion search experiments, cylindrical microwave cavities are typically employed to detect signals from the axion-photon conversion.
To enhance the conversion power and reduce the noise level, cavities are placed in strong solenoid magnetic fields at sufficiently low temperatures.
Exploring high mass regions in cavity-based axion search experiments requires high frequency microwave cavities and thus understanding cavity properties at high frequencies in extreme conditions is deemed necessary. 
We present a study of the magnetoresistance of copper using a cavity with a resonant frequency of 12.9 GHz at the liquid helium temperature in magnetic fields up to 15\,T utilizing a second generation high temperature superconducting magnet. 
The observations are interpreted to be consistent with the anomalous skin effect and size effect.
This is the first measurement of magnetoresistance at a high frequency ($>10$\,GHz) in high magnetic fields ($>10$\,T).
}

\keywords{Dark Matter detectors (WIMPs, axions, etc.); Detector design and construction tech- nologies and materials; Microwave radiometers}

\arxivnumber{1705.04754}

\begin{document}
\maketitle
\flushbottom

\section{Introduction}
The axion, a hypothetical elementary particle postulated by R. Peccei and H. Quinn to solve the “strong-CP” problem in quantum chromodynamics (QCD) of particle physics, is the result of a breakdown of Peccei-Quinn (PQ) global symmetry~\cite{bib:PQ}. 
This particle has been considered as an attractive candidate for cold dark matter with its mass falling within a specific range. Typical experimental approaches for axion search adopt the methodological concept suggested by P. Skivie of placing microwave resonant cavities in strong magnetic fields, where axions are converted into microwave photons ~\cite{bib:Sikivie}.
The conversion power of the axion signal is expressed by 
\begin{equation}
P_{a\rightarrow\gamma\gamma} = g_{a\gamma\gamma}^2 \frac{\rho_a}{m_a}B^2VC\,  {\rm{min}}(Q_L,Q_a),
\label{eq:power}
\end{equation}
where $g_{a\gamma\gamma}$ is the axion-photon coupling constant, $\rho_a$ is the axion local density, $m_a$ is the axion mass, $B$ is the external magnetic field, $V$ is the cavity volume, $C$ is the mode-dependent form factor, and $Q_L$ and $Q_a$ are the cavity and axion quality factors, respectively. 
An important quantity relevant to the experimental sensitivity is the signal-to-noise ratio (SNR), which is given by
\begin{equation}
{\rm{SNR}}\equiv \frac{P_{\rm{signal}}}{P_{\rm{noise}}} = \frac{P_{a\rightarrow\gamma\gamma}}{k_BT_{\rm{syst}}} \sqrt{\frac{t_{\rm{int}}}{\Delta f_a}},
\label{eq:snr}
\end{equation}
where $k_B$ is the Boltzmann constant, $T_{\rm{syst}}$ is the total system temperature, $t_{\rm{int}}$ is the integration time, and $\Delta f_a$ is the signal bandwidth.
As can be seen in Eqs.~(\ref{eq:power}) and~(\ref{eq:snr}), $Q$, along with $B$ and $T$, is an important experimental parameter for enhancing the sensitivity.

The mass of the halo dark matter axion, if it exists, is believed to lie between 1$\,\mu$eV and 1$\,$meV.
Under the assumption that the PQ symmetry broke after inflation, theoretical calculations prefer a high mass axion, for instance, $m_a \gtrsim 50\,\rm{\mu} eV$ (corresponding frequency $f_a \gtrsim 12\,\rm{GHz}$) according to recent reports~\cite{bib:nature}\cite{bib:update}.
This mass range is at least 2 times higher than those reachable by the resonant microwave cavity experiments currently in operational mode~\cite{bib:ADMX}\cite{bib:ADMX_HF}.
Since the signal frequency is proportional to the axion mass, searching for higher mass axions requires higher frequency microwave cavities.
In this context, therefore, understanding physical properties of RF cavities at high frequencies in strong magnetic fields at low temperatures is of importance.

The quality factor of a cavity is directly related to the surface resistance, $R_s$, as in
\begin{equation}
Q=\frac{\Lambda}{R_s},
\label{eq:Q_R}
\end{equation}
where $\Lambda$ is a constant that can be calculated depending on the cavity geometry and the resonant mode.
Magnetoresistance, the tendency of the electrical resistance of a material to vary in an external magnetic field, 
is generally expressed as the fractional change in surface resistance due to the field.
It is equivalent to the negative of the fractional change in the quality factor from Eq.~(\ref{eq:Q_R}), forming a relation
\begin{equation} 
\frac{\Delta R_s}{R_s} = - \frac{\Delta Q}{Q}.
\label{eq:magnetoresistance}
\end{equation}
It should be noted that our interest is in the normalized value of a quantity rather than its absolute value, and thus various systematic uncertainties are cancelled out for the measurement.

Since its first observation in 1940~\cite{bib:london}, the anomalous behaviour of the electrical resistivity in metals in extreme conditions (at low temperatures and high frequencies) has been understood as resulting from the anomalous skin effect~\cite{bib:pippard}. 
The classical skin effect, in the limit of $f \ll 1/\rho\epsilon$, describes the skin depth explicitly as
\begin{equation}
\delta=\sqrt{\frac{2\rho}{\mu\omega}},
\label{eq:skin_depth}
\end{equation}
where $\rho$, $\epsilon$, and $\mu$ are the resistivity, permittivity, and permeability of the metal and $\omega$ is the angular frequency of resonant mode under consideration. 
For a 1$\,$GHz copper cavity with $\rho=1.7 \times 10^{-8}\,\Omega$m, the skin depth is 2.1$\,\mu$m at room temperature.
This skin effect limits the conductivity of metals according to
\begin{equation}
\rho = R_s\cdot\delta.
\label{eq:R_delta}
\end{equation}

It is also known that a magnetic field induces non-zero change in surface resistance: in general, the resistivity of metals increases with field~\cite{bib:olsen}.
However, an abnormal decrease of electrical resistivity is also observed in metal wires or thin films in the presence of magnetic fields.
This abnormal behaviour results from the cyclotron motion of conduction electrons induced by the Lorentz force.
In semi-classical approaches, the average radius of the cyclotron orbit under magnetic field $B$ is expressed as,
\begin{equation}
r_c  = \frac{\sqrt{2m_eE_F}}{eB},
\label{eq:cyclotron_radius}
\end{equation}
where $m_e$ is the electron mass,  $E_F$ is the Fermi energy of the metal, and $e$ is the electron charge.
With increasing magnetic field, the conduction electrons are confined in smaller orbits resulting in fewer collisions with the metal surface; thus, surface resistivity becomes abnormally small until the normal behaviour becomes dominant~\cite{bib:size_effect}.
For copper, using $E_F=7.0\,$eV, the average cyclotron radius is $8.9\,\mu$m under a 1$\,$T field.

Such anomalous behaviours of electrons give rise to asymptotic saturation of resistivity in metals at low temperatures and in high magnetic fields.
This corresponds to the size effect observed in thin metals, which occurs when the skin depth and cyclotron radius are roughly the same as the electron mean free path
\begin{equation}
\lambda  = \frac{\sigma m_e v_F}{ne^2},
\label{eq:mean_free_path}
\end{equation}
where $\sigma$ and $n$ are the conductivity and the free electron density of the metal and $v_F$ is the Fermi speed.
As the resistivity decreases so does the skin depth, while the mean free path increases and eventually reaches the same order as the skin depth, where the resistivity is minimized and  saturation takes place.
Using a Fermi speed of about $1.6\times10^6\,$m/s, the mean free path of an electron in copper can be calculated and is found to be $0.04\,\mu$m.

A previous measurement of the surface resistance using a 1.25$\,$GHz frequency copper cavity in a 5.8$\,$T magnet revealed that the AC magnetoresistance is an order of magnitude smaller than the DC magnetoresistance and depends quadratically on the field~\cite{bib:previous}. 
In this paper, we report a study of magnetoresistance in copper at a higher frequency of 12.9$\,$GHz in higher magnetic fields up to 15$\,$T at the liquid helium (LHe) temperature of 4.2$\,$K.

\section{Experimental Setup}
One of the major components of the experiment is a high temperature superconducting (HTS) magnet, designed by the Francis Bitter Magnet Laboratory at MIT and constructed and tested by SuNAM Co., Ltd~\cite{bib:sunam}.
This magnet features the multi-width (MW) design consisting of 26 double-pancake coils with 5 various widths wound up with second generation (2G) GdBCO HTS tapes, which allows for a higher overall current density for the same operating current~\cite{bib:hahn2}.
It is also incorporated with no-insulation (NI) winding technique enabling HTS magnets to be self-protecting against potentially destructive quenches~\cite{bib:hahn1}.
The highly compact design generated a center field of 26.4\,T at an operating current of 242\,A in a LHe bath, a record high in magnetic fields from all HTS magnets.
For this experiment, only a current of 140\,A is applied because one of the double-pancake coils shows an abnormally increasing terminal voltage at over 150\,A.

The magnet has a 25 (35)\,mm clear (winding) bore and a field uniformity of 99\% within $\pm\,25$\,mm from the field center along the axial direction and within $\pm\,9$\,mm along the radial direction, which determines the dimensions of the cavity for the experiment. 
The magnet is immersed in the LHe inside a cryostat with 250\,mm inner diameter and 1140\,mm depth. 
A pair of vapour-cooled current leads are penetrated into the cryostat and connected to the magnet. 
The magnet is at 35\,mm above the bottom of the cryostat and seven radiation shielding disks approximately 250\,mm in diameter are placed between the magnet and the cryostat cap to improve the thermal insulation.
Two temperature sensors are mounted on the top and bottom of the magnet to monitor the magnet temperature during the experiment.

A cylindrical microwave resonant cavity with an 18\,mm inner diameter and 50\,mm inner height is exploited in this experiment.
This cavity is made of oxygen-free high conductivity copper (OFHC) with 99.99\% nominal purity.
The cavity dimensions are determined by the magnet bore size and the field distribution, such that the corresponding TM$_{010}$ resonant frequency is 12.8 GHz at room temperature and the averaged field is 99\% of the maximum field.
The cavity consists of two identical pieces of vertically cut half-cylinder, which are hollowed out, and tied with copper flanges.
The electrical field generated in the TM$_{010}$ mode is parallel to the cavity axis and bounced by the end caps to induce longitudinal surface current along the wall. 
In conventional cavity design, in which inevitable mechanical gaps between the end caps and the cavity wall are introduced, the surface current may experience contact resistance when flowing across gaps.
The concept of this split cavity is adopted to eliminate such discontinuity in the current path between cavity wall and end-caps .
The cavity is fabricated using a machining center tooling system and the surface is highly polished using compounds with particle size down to $4\,\mu$m .

\begin{figure}[h]
\centering
\includegraphics[width=0.55\textwidth]{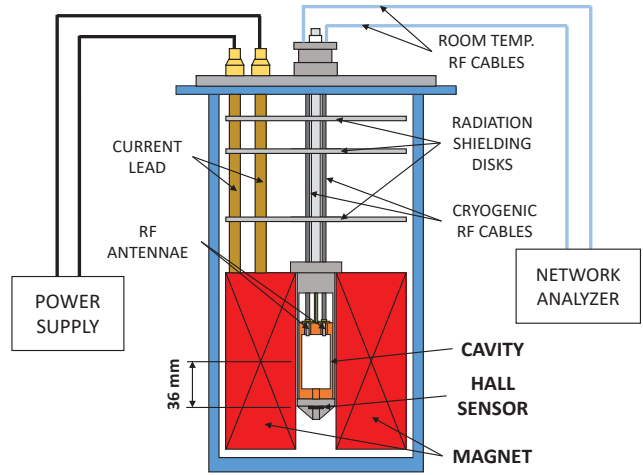}
\caption{Schematic view of the experimental setup. 
The major components of the system are a copper cavity (orange) with a Hall sensor (black) mounted on the bottom and a HTS high field magnet (red) immersed in LHe within a cryostat (blue). 
The quality factor is measured using a network analyser through a pair of weekly coupled RF antennae.}
\label{fig:setup}
\end{figure}

So that it can be stably positioned at the center of the magnet, the cavity is sustained by a support structure made out of stainless steel 316L.
A pair of RF antennae are weakly coupled with the cavity to measure the unloaded quality factor.
Signals are delivered by a pair of cryogenic RF cables to the feedthrough on the top flange of the cryostat and through a pair of room temperature cables to a network analyzer.
A Hall sensor is mounted on the bottom of the support structure 36 mm below the center of the magnet.
The overall structure of the experimental setup is shown in Fig.~\ref{fig:setup}.

\section{Measurements}
Due to the spatial limitation of the magnet bore, only the TM$_{010}$ mode is considered for this measurement.
The measured resonant frequency at room temperature is 12.82$\,$GHz, consistent with the estimated value within a typical machining tolerance of 50$\,\mu$m.
The quality factor, $Q_m$, is measured using the 3 dB method through transmission signal between the pair of RF antennae, which are weakly coupled ($\sim0.1\,$dB) with the cavity.
The coupling strengths of these antennae are monitored over the exercise and accounted for to obtain the unloaded quality factor, $Q_0$, using the following relation
\begin{equation} 
Q_0 = Q_m(1+\beta_1+\beta_2),
\label{eq:unloaded_Q}
\end{equation}
where $\beta_i=Q_0/Q_i$ is the coupling coefficient of antenna $i$, which is determined by the reflection coefficient $\Gamma$ and scattering parameter $S$ as
\begin{equation} 
\Gamma_i = \left|\frac{1-\beta_i}{1+\beta_i}\right| \,\, {\rm{and}} \,\, S_{ii}=-{\rm{log}} \Gamma_i^2.
\end{equation}

The resonant frequency drops to 12.56$\,$GHz when the cavity is filled up with LHe.
Taking the dielectric constant of the cryogen ($\epsilon_r=1.055$) into account, the corresponding value in vacuum is estimated to be 12.90$\,$GHz.
A slight increase in frequency is attributed to the thermal shrink and is consistent with expectations.
The coupling strength of the antennae changes up to 0.4 dB at the LHe temperature.
The corrected resonant frequencies and unloaded quality factors of the TM$_{010}$ mode at 300$\,$K and 4.2$\,$K are summarized in Table~\ref{tab:Q_f}.
\begin{table}[hbt]
\centering
\caption{Resonant frequencies and unloaded quality factors of the TM$_{010}$ mode at 300$\,$K and 4.2$\,$K and zero magnetic field. The dielectric constant of LHe and the effect of antenna coupling are taken into account. The errors include the statistical uncertainties. The uncertainties of the frequency measurement are negligible.}
\label{tab:Q_f}
\begin{tabular*}{0.475\textwidth}{@{\extracolsep{\fill}}ccc}
\toprule
&$f$ [GHz]&$Q_0$\\
\midrule
300\,K & 12.82 $\pm$ 0.00 & 12,160 $\pm$ 79  \\
\,4.2\,K & 12.90 $\pm$ 0.00 & 33,686 $\pm$ 84 \\
\bottomrule
\end{tabular*}
\end{table}

\begin{figure}[h]
\centering
\includegraphics[width=0.48\textwidth]{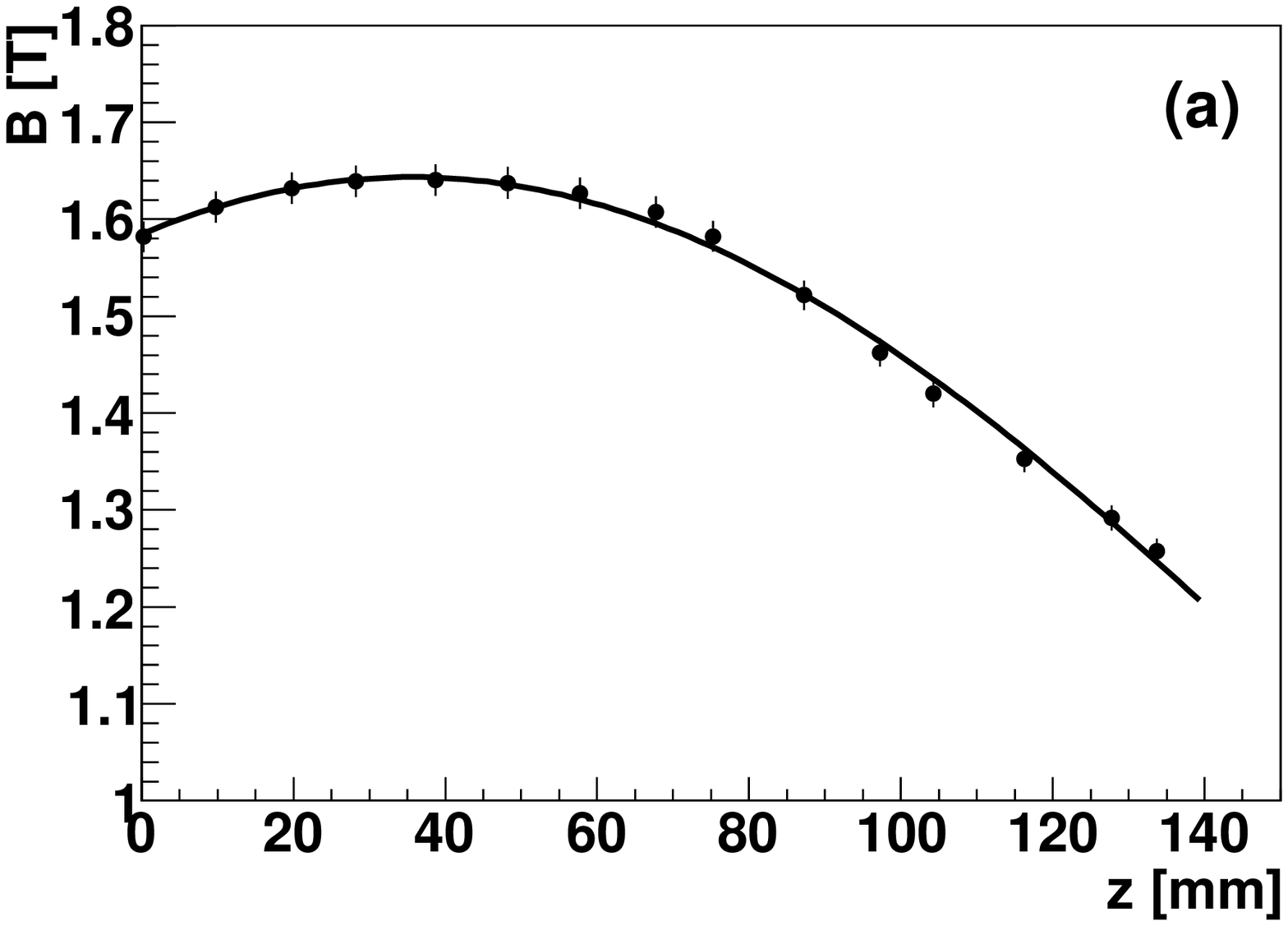}
\includegraphics[width=0.48\textwidth]{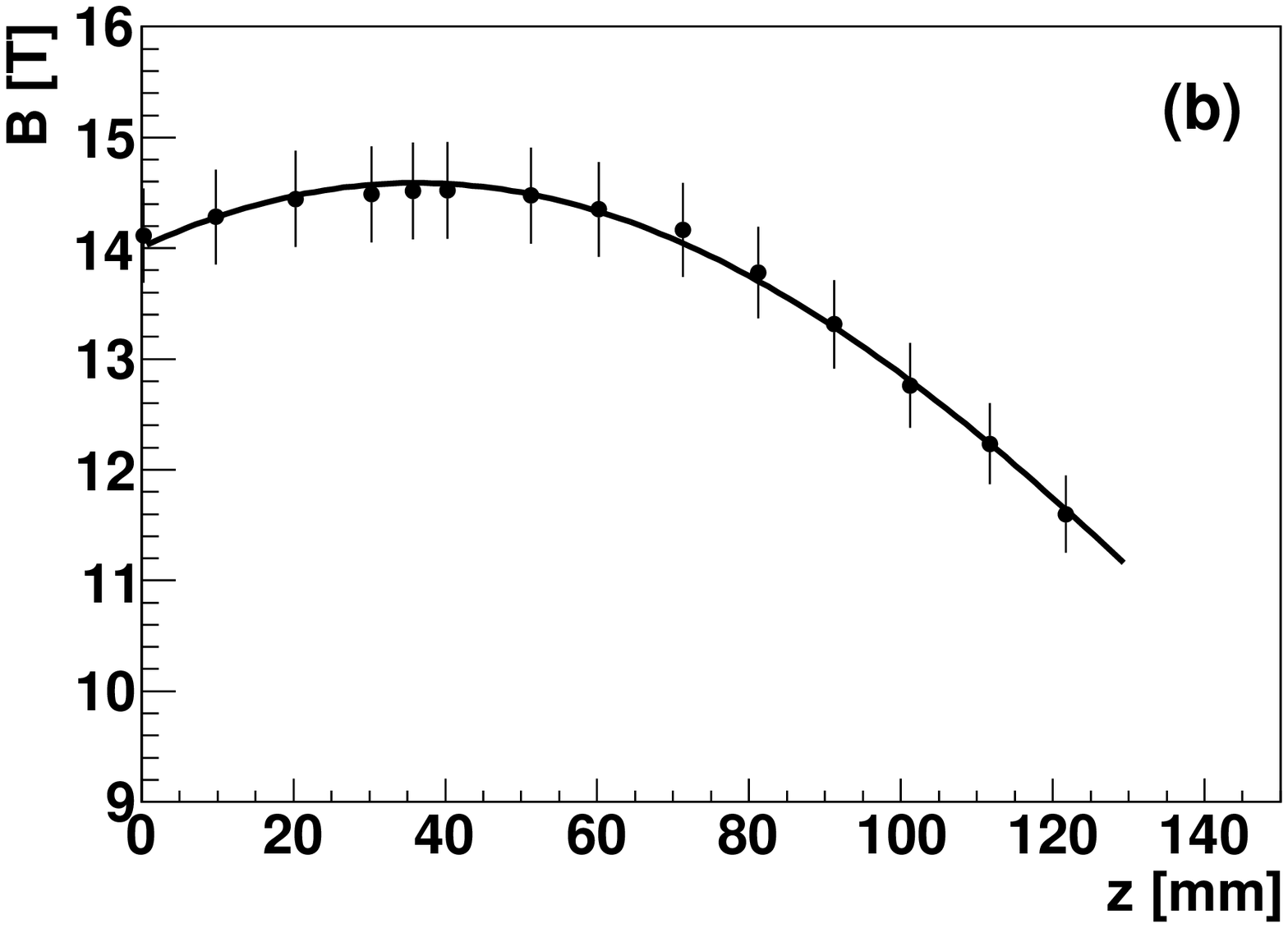}
\caption{Field distributions along the magnet axis at the center of the bore at a constant power supply current of 15$\,$A in 77$\,$K LN$_2$ (a) and at 140$\,$A in 4.2$\,$K LHe (b).
The error bar of each data point represents the systematic uncertainty of measurement by the Hall sensor. The statistical uncertainty is negligible. 
The fitting procedure assumes that the data follow a Gaussian distribution.
Note the Hall sensor is located 36\,mm below the center of the magnet.}
\label{fig:field_map}
\end{figure}

The magnet is initially tested at 77$\,$K in liquid nitrogen (LN$_2$).
At a constant power supply current of 15$\,$A, a field map is drawn along the axial direction.
The distribution is fitted using a Gaussian distribution, from which we obtain a maximum $B$ field of 1.64$\,$T and a field center of $35.6\pm2.2\,$mm above the Hall sensor.
The distribution shows good agreement with the simulated distribution provided by SuNAM and the obtained value is consistent with the physical location of the sensor with respect to the magnet center.
An additional field map is drawn at a current of 140$\,$A when the magnet is energized at 4.2$\,$K in LHe. 
The same fitting procedure to the distribution returns a maximum $B$ field of 14.6$\,$T at a relative sensor position of $36.2\pm6.3\,$mm.
The consistency between the measured and expected field distributions and relative positions of the Hall sensor confirms the performance of the magnet.
The measured field distributions at 15$\,$A and 140$\,$A are shown in Fig.~\ref{fig:field_map}.
The average difference in the field strength between the cavity center and the Hall sensor position is 3.9\% and taken into account when estimating the field at the cavity center.

The $Q$ measurement is performed while the magnet is energized by a power supply at a ramping rate of 15\,mA/s{\footnotemark} in 4.2\,K LHe.
Every 2 seconds, the cavity quality factor is measured by the RF antennae through the transmission signal and the magnetic field is recorded by the Hall sensor at its position.
The coupling strengths of the antennae are taken into account to obtain the unloaded $Q_0$ of the cavity using Eq.~(\ref{eq:unloaded_Q}).
The averaged magnetic field within the cavity (99\% of the maximum field) is used and a scale factor of 1.039 is applied to account for the offset of the Hall sensor position with respect to the magnet center. 
At 140\,A, the Hall sensor reads a field strength of 14.1\,T, at which the aforementioned field map is drawn.
The measurement is repeated for consistency check while the magnet is de-energized at the same ramping rate.
\footnotetext{At this ramping rate, the flux change owing to the eddy current is negligible.}

\section{Results}
\begin{figure}[h]
\centering
\includegraphics[width=0.48\textwidth]{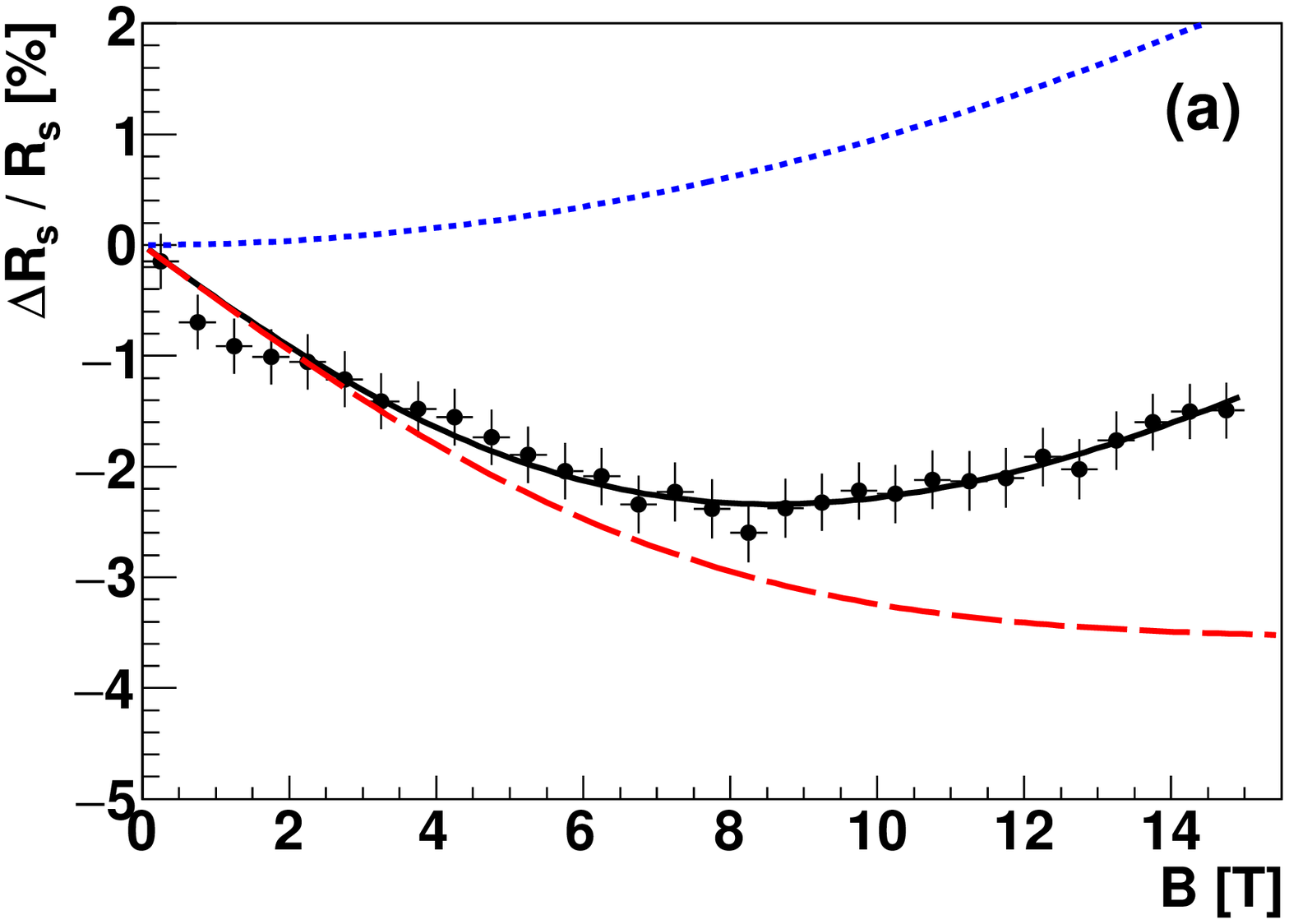}
\includegraphics[width=0.48\textwidth]{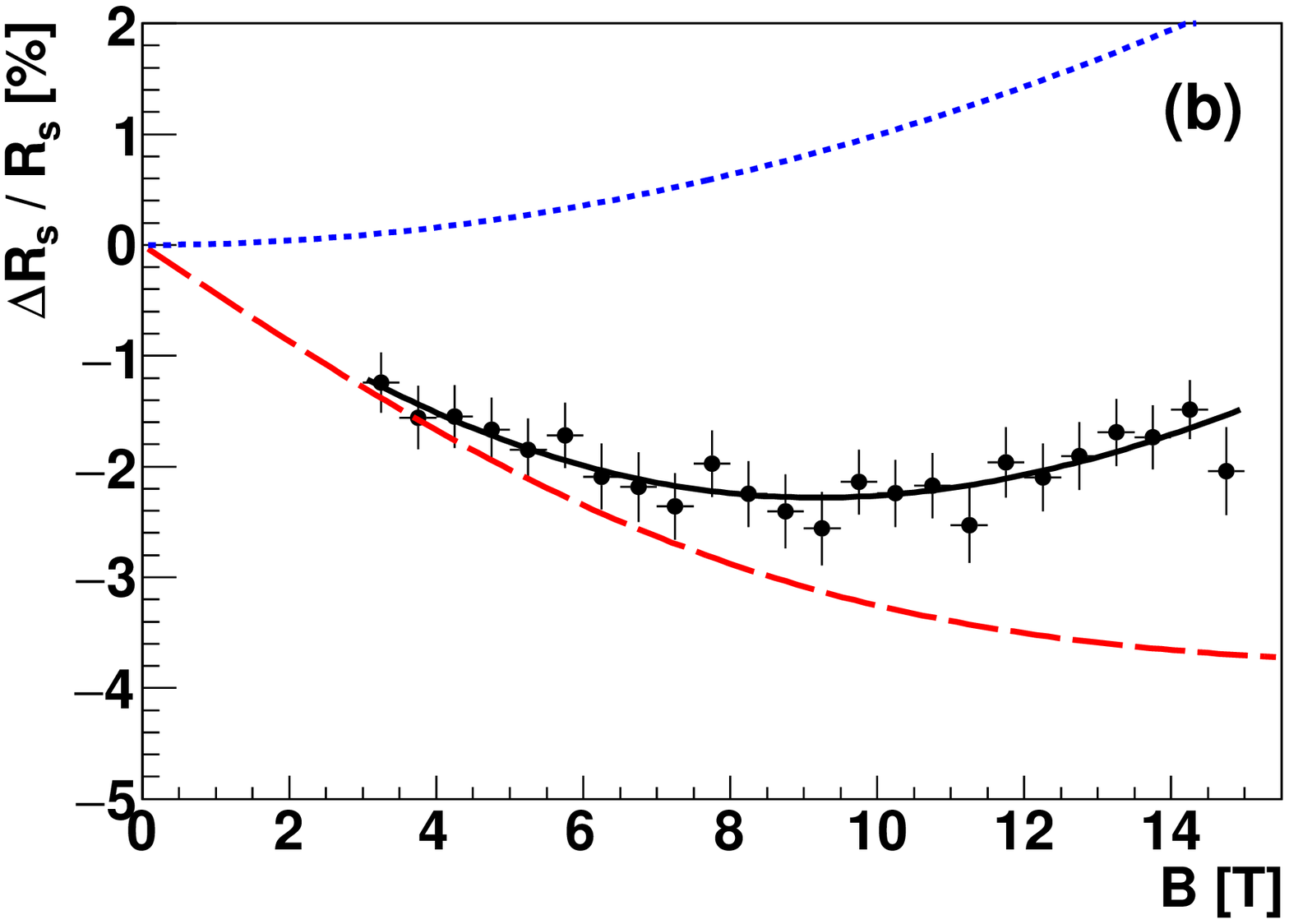}
\caption{Fractional change in surface resistance while charging (a) and discharging (b) the magnet.\protect\footnotemark[2]
The error bar of each data point includes the statistical uncertainty of $Q$ value measurement.
The distributions are modelled by a linear combination of the second order polynomial with the two lowest degree coefficients fixed to zero, represented by the dotted lines in blue, to describe the normal behaviour of the magnetoresistance and the Error function, represented by the dashed lines in red, to describe the abnormal behaviour.}
\label{fig:magnetoresistance}
\end{figure}

Magnetoresistance is represented by the fractional change in electrical resistivity of a material under a magnetic field.
Using Eq.~(\ref{eq:magnetoresistance}), we transform the unloaded cavity quality factors into the electrical surface resistivity. 
Fig.~\ref{fig:magnetoresistance} shows the fractional change in $R_s$ as a function of $B$ field over the charging cycle of the magnet.
A parabolic shape with a negative minimum value is commonly observed.
\footnotetext[2]{A truncation at around 3$\,$T in plot (b) is due to temperature rise in the magnet owing to failure of refilling the cryotank with LHe in a timely manner while discharging the magnet.}

We assume that the total magnetoresistance can be decomposed into two components: the normal component, for which the resistance in single copper crystal increases quadratically in $B$~\cite{bib:olsen}, and the abnormal component, for which the resistance decreases with field similar to the decrease of the resistance with temperature.
The abnormal decreases in surface resistivity at low fields are explained by the relative size difference between the cyclotron radius and the skin depth of the metal.
Free electrons in spiral motion in the presence of an external magnetic field experience resistance from collisions with the metal surface.
Such resistance decreases, with the increasing of external field, until the size of the cyclotron motion reaches the same order as the skin depth size within which the electrons flow.
Beyond that point, the abnormal component remains unchanged while the normal behaviour of the magnetoresistance becomes dominant, i.e., electrical resistance increases with field.
This describes the parabolic behaviour of the surface resistance with negative change at relatively low magnetic fields.
Based on this assumption, we model the normal contribution with a quadratic function and the abnormal contribution with an Error function and obtain a maximal changes of surface resistance of $2.3\pm0.8\%$ at $8.6\pm3.1\,$T while charging the magnet and $2.3\pm1.7\%$ at $9.2\pm6.3\,$T while discharging the magnet.
We also find that the abnormal components become saturated at $15.0\pm3.2\,$T and $17.8\pm5.8$\,T and that their maximal contributions are $3.5\pm0.8\%$ and $3.8\pm1.8\%$, respectively.

An interpretation of the abnormal behaviour can be made as follows.
From the relationship between the quality factor and the skin depth, as given in Eqs.~(\ref{eq:Q_R}), (\ref{eq:skin_depth}), and (\ref{eq:R_delta}), the residual resistivity ratio, RRR, can be obtained as $\rho_{300{\rm{K}}} /\rho_{4{\rm{K}}} = (Q_{4{\rm{K}}}/Q_{300{\rm{K}}})^2=7.7$.
Taking this value into account, the classical skin depth for our cavity and the expected mean free path at 4.2$\,$K are obtained as $\delta = 0.21\,\mu$m and $\lambda= 0.30\,\mu$m from Eq.~(\ref{eq:skin_depth}) and Eq.~(\ref{eq:mean_free_path}), respectively.
The radius of the cyclotron orbit of conduction electrons in copper at the averaged saturation field of 15.7$\,$T field is calculated from the semi-classical approach, using Eq.~(\ref{eq:cyclotron_radius}), and is found to be $r_c=0.57\,\mu$m.
It should be noted that these three quantities are of the same order.
Similar to the anomalous skin effect, the magnetoresistance is expected to be minimized when conditions in which the three quantities are compatible in size are realized.

The previous measurement, in which the frequencies under consideration are around 1.5\,GHz, observed similar parabolic behaviour of the magnetoresistance of the different modes.
In particular, it is seen that a minimum of $\Delta R_s/R_s$ takes place at around 3\,T and a saturation of the residual $\Delta R_s/R_s$ is reached at around 5$\,$T for the TM$_{010}$ mode.
Since $\delta\sim1/\sqrt f$, from Eq.~(\ref{eq:skin_depth}), our skin depth is to be roughly 3 times smaller when the cavity frequency is about 10 times higher.
Therefore, it is expected that a magnetic field that is stronger by a similar factor will be necessary for the size of the cyclotron orbit to be compatible and for a saturation in surface resistance to appear for larger fields, which are observed in our measurements.

\section{Conclusions}
In summary, we studied the copper magnetoresistance at high frequency as a function of magnetic field by measuring the quality factor $Q_0$ of a 12.9$\,$GHz copper cavity with gradually increasing fields up to 15\,T at 4.2\,K.
A quadratic dependence of the magnetoresistance on the field with a negative minimum value of about a few percent, taking place at around 9\,T, is observed.
This behaviour is well explained by the geometric size effect, i.e., the resistance decreases until the average cyclotron radius of the conduction electrons becomes compatible with the mean free path and the skin depth.
The resulting behaviour of the quality factor, varying by about a few percent over a large field range, indicates that the influence of strong magnet fields on the cavity-based axion search experiments exploring high mass regions will be tolerable.
It is also desirable to repeat the measurement at different frequencies to generalize the anomalous behaviour of the magnetoresistance as a function of the $B$ field.
Finally, this measurement is the first user-based physics result employing the MW NI 2G HTS magnet technologies for the scientific purpose.
Finally, this measurement is the first application of MW NI 2G HTS magnet technologies to a scientific study.

\acknowledgments
This work was supported by IBS-R017-D1-2017-a00 / IBS-R017-G1-2017-a00 / IBS-R017-Y1-2017-a00. 
We thank Jaemin Kim, Kanghwan Shin, and Sangwon Yoon from SuNAM Co., Ltd. for providing the 26\,T HTS magnet and assisting in the operation.

\end{document}